\documentclass[%
rmp,reprint,letterpaper,%
superscriptaddress,citeautoscript,%
footinbib]{revtex4-1}
\usepackage[usenames,dvipsnames]{xcolor}
\usepackage[bookmarks=false,colorlinks]{hyperref}
\hypersetup{
    linkcolor=Red,       
    urlcolor=NavyBlue, 
    citecolor=ForestGreen,     
}
\usepackage{amssymb}
\usepackage{amsmath}
\usepackage{graphicx}
\usepackage{microtype}

\begin{document}

\title{Emergent properties hidden in plane view: Strong electronic correlations at oxide interfaces}
\author{Jak Chakhalian}
\affiliation{Department of Physics, University of Arkansas, Fayetteville, AR 72701, USA}%

\author{John W.\ Freeland}
\affiliation{Advanced Photon Source, Argonne National Laboratory, Argonne, Illinois 60439, USA}%

\author{Andrew J.\ Millis}
\affiliation{Department of Physics, Columbia University, 538 West 120th Street, New York, New York 10027, USA}%

\author{Christos Panagopoulos}
\affiliation{School of Physical and Mathematical Sciences, Division of Physics and Applied Physics, Nanyang Technological University, 637371 Singapore}%

\author{James M.\ Rondinelli}
\altaffiliation[Present Address: ]{Department of Materials Science \& Engineering, Northwestern University, Evanston, Illinois, 60208, USA}%
\affiliation{Department of Materials Science \& Engineering, Drexel University, Philadelphia, Pennsylvania 19104, USA}

\begin{abstract}
Complex oxides with correlated carriers are a class of materials characterized by multiple competing and nearly degenerate ground states due to interactions that create a subtle balance to define their ground state. This in turn leads to a wide diversity of intriguing properties ranging from high Tc superconductivity to exotic magnetism and orbital phenomena. By utilizing bulk properties of these materials as a starting point, interfaces between different classes of correlated oxides offer a unique opportunity to break the fundamental symmetries of  the bulk and alter the local environment. From experimental  point of view, utilizing recent advances in  growth with atomic layer precision one can now combine layers of compounds   with distinct and even antagonistic order parameters to design new artificial  quantum materials. Here we illustrate this approach by selected examples how  broken lattice symmetry, strain, and altered chemical and electronic environments at the correlated interfaces can provide a unique platform to manipulate this subtle balance and enable  quantum many-body states and phenomena not attainable in the bulk. 
\end{abstract}

\maketitle

\tableofcontents

\section{Introduction}
Finding new collective electronic states in materials is one of the fundamental goals of  condensed matter physics.  While the traditional approach has been to search for such phases within naturally occurring compounds, in recent years the focus has shifted to \emph{heterostructures} \cite{Hwang:2012p38867}: artificial materials formed by interleaving two or more structurally and chemically dissimilar materials. Of particular interest is the spatial region at the  \emph{interface} where dissimilar  materials meet. New states may emerge here because the environment near an interface is different from that occurring in bulk (thermodynamically stable) materials.
Advances in the \emph{angstrom-scale}  layer-by-layer synthesis of multi-element compounds for \emph{materials-by-design} have taken the approach to a new level of power and sophistication: It enables the atomic-scale combination of materials with different properties, granting  access to a new  terrain in which unusual  states of matter may
arise \cite{Schlom:2008p15406}.

Heterostructures formed from transition metal oxides (TMO) are a particularly appealing hunting ground for new physics. In these materials the transition metal ($M$) ion has an open $d$-shell  electronic configuration with spin, orbital, and charge degrees of freedom. Electrons in these partially filled $d$-shells are \emph{correlated}:  the motion of one electron depends explicitly and non-trivially on the behavior of all of the others giving rise to  interesting many-body phenomena \cite{Imada:1998p38738}. The resulting magnetic, superconducting, and multiferroic phases are of great scientific interest and are potentially capable of providing innovative energy, security, electronics and medical technology platforms. The heterostructure geometry  \cite{Zubko:2011p42313,Mannhart:2010p14504,MRS:9125042} enables otherwise  unattainable changes in atomic structure and chemical bonding, leading to new modalities for control and optimization of known states and potentially leading to new ones.

Over the past decade, one particular class of heterostructures, based on the interface between lanthanum aluminate (LaAlO$_3$; LAO for short) and Strontium titanate (SrTiO$_3$; STO for short), has been the subject of very extensive study. In this Colloquium we choose not to discuss the LAO/STO interface or its variants, selecting our examples instead from  vanadate, manganite, cuprate and nickelate-based systems for two reasons.  First, the LAO/STO system and its variants have been extensively reviewed in other venues, see for example \onlinecite{Mannhart08,Zubko11,Hwang:2012p38867}.
Second, and more importantly, the LAO/STO system involves doping nominally insulating STO with maximum sheet carrier densities of fewer than 0.5 electrons ($e$) per in-plane unit cell, and the charge density is typically spread over several unit cells in the direction perpendicular to the interface. The volume carrier densities are therefore typically low, so that the situation is more closely related to a doped semiconductor than to the correlated electron materials on which we wish to focus here.
Nonetheless some of the scientific excitement has focused on superconductivity \cite{Reyren:2007p42482} (exhibited also when bulk STO is lightly doped by conventional dopants) and Anderson localization metal-insulator transitions  \cite{Caviglia08}.
Reports or theoretical suggestions of other correlation phenomena in this system including charge ordering \cite{Pentcheva07} and magnetism \cite{Okamoto:2006p42492,Brinkman07,Ariando11,Bert11,Li11}  have also appeared; we return to the issue in section IV and in the conclusions.

\begin{figure*}[t]
   \centering
  \includegraphics[width=1.11\columnwidth]{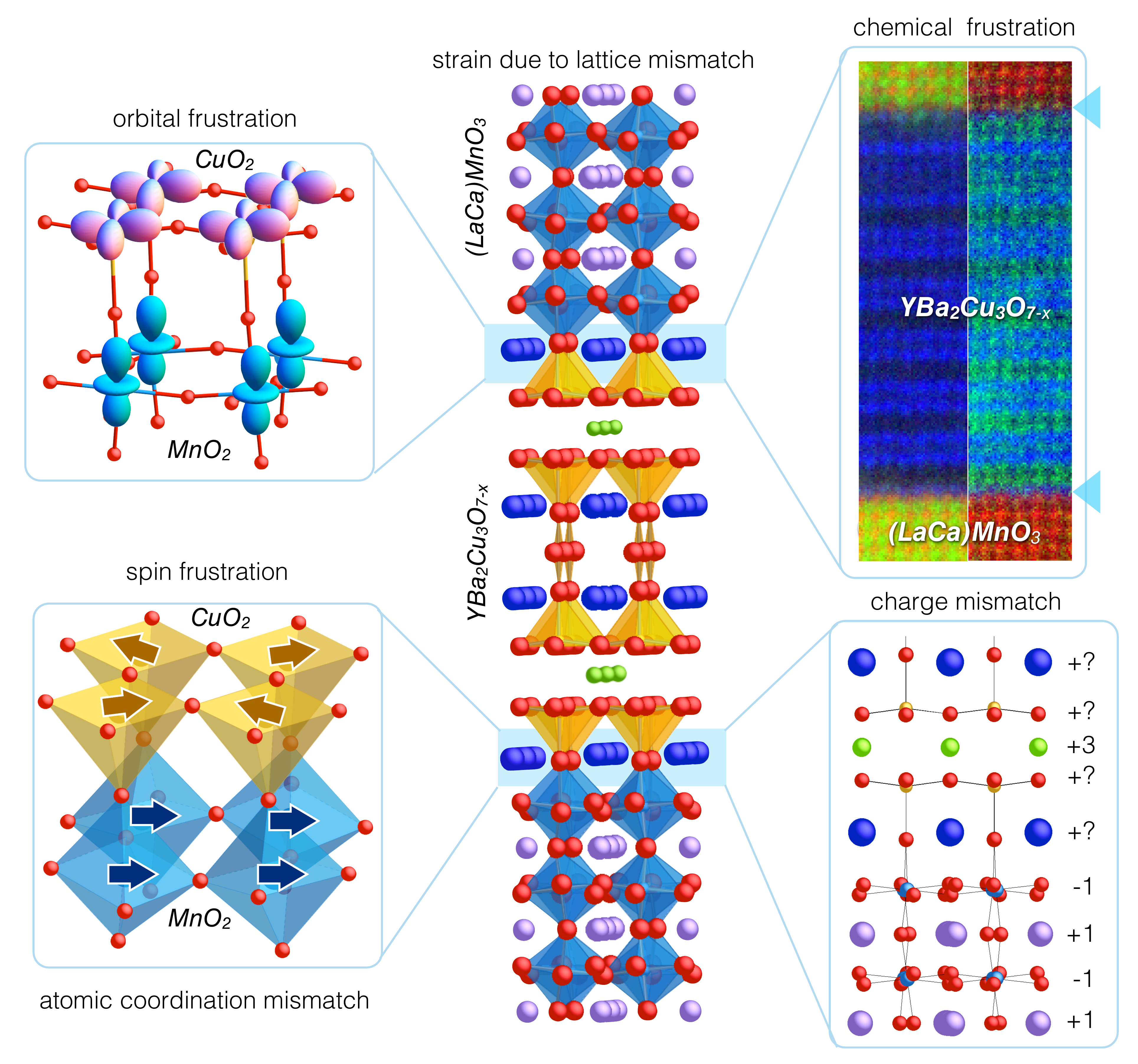}\vspace{-0.4\baselineskip}
   \caption{Anatomy of an oxide heterointerface: an illustration showing the interplay between different degrees of freedom (charge, spin, and lattice) at a coherently grown interface between ferromagnetic La$_{2/3}$Ca$_{1/3}$MnO$_3$ and superconducting YBa$_2$Cu$_3$O$_{7-x}$. The 
electron micrograph is reproduced from Ref.~\onlinecite{Chien:2013p42266}, where each color represents a different chemical species.
   \label{fig:hetero_anatomy}}
\end{figure*}

This Colloquium  illustrates the essential features that make TMO-based heterostructures an appealing discovery platform for emergent properties. The guiding principle is that strong electronic correlations in combination with the access to new symmetries and electronic band structures provided by  oxide interfaces can activate new electronic properties formerly ``hidden'' in bulk compounds. We illustrate this principle with a few selected examples, showing how charge redistributes, magnetism and orbital polarization arises and ferroelectric order emerges from heterostructures comprised of oxide components with nominally contradictory behavior. For example, interfaces may be metallic, magnetic, or ferroelectric even though in bulk form the constituent  materials are insulating, non-magnetic, or simple dielectrics. We conclude by articulating open challenges and opportunities in the field, in particular, how to translate the new understanding of when emergent phases arise into control of novel behavior by design at oxide interfaces, and the manipulation of these states by suitable mechanical, electrical or optical  boundary conditions and excitations.

\section{Anatomy of an Oxide Interface}
The formation of a coherent perovskite oxide heterointerface, as shown
in \autoref{fig:hetero_anatomy}, provides a remarkable correlated electron ``playground.'' It brings different transition metal cations with their localized $d$ electron physics and interacting charge, spin and lattice degrees of freedom into intimate contact in a tunable crystalline environment.
The key structural features  of transition metal oxides relate to the coordination geometry of the metal ions and the metal-oxygen-metal bond angles. These determine  magnetic exchange interactions~\cite{Kanamori:1965p42321,Goodenough:1955p42314,Anderson:1950p42315} and electronic bandwidths~\cite{Eng:2003p42319}, thereby controlling the electronic and magnetic ground states. Structural and electronic changes across an interface can act to  stabilize previously unanticipated phases of matter \cite{Okamoto:2004p42312}.
Consider for  example a multilayer heterostructure comprised of alternating blocks of the metallic ferromagnet La$_{2/3}$Ca$_{1/3}$MnO$_3$ (LCMO), and the high-temperature cuprate superconductor YBa$_2$Cu$_3$O$_{7-x}$  (YBCO) sketched in \autoref{fig:hetero_anatomy}.  The interface brings several crucial structural effects. The first is a coordination mismatch. LCMO is a three-dimensional perovskite ($AM$O$_3$ stoichiometry) with corner-connected MnO$_6$ octahedra that may be described by interleaving alternating (La,Ca)O and MnO$_2$ layers along [001]. In contrast, YBCO is a two-dimensional oxide with four- and five-fold coordinated Cu cations. The layered cuprate structure may be considered as a derivative of
perovskite, which partly facilitates coherent growth of the heterostructure. But unlike LCMO, YBCO displays an ordered network of oxygen vacancies accommodated by the valence preferences of Cu: One oxygen atom is removed from every third (001) YO plane to produce the square pyramidal CuO$_5$ coordination, then on every third CuO$_2$ layer, vacancies order along [100], producing the square planar CuO$_4$  coordination. Thus a  ``coordination mismatch''
arising from the change from the 6-fold coordination of the Mn to the lower coordination of the Cu (\autoref{fig:hetero_anatomy}, lower left) occurs at the interface. As a result, a set of CuO chains ($i.e.$ charge reservoir) is missing from  the interfacial YBCO unit cell to maintain a prerovskite-like  sequence ...MnO$_2$--BaO--CuO$_2$... across the junction~\cite{Zhang:2009p13098,Chien:2013p42266}.

Coherent epitaxial  growth also produces an intrinsic  strain mismatch arising from the  different equilibrium lattice constants (\autoref{fig:hetero_anatomy}, center). The atomic structure at the heterointerface responds to alleviate the strain mismatch through relaxation of the interatomic distances and internal atomic degrees of freedom (for example, rotations or size deformations to the transition metal oxygen polyhedra)  in the constituents along the  superlattice repeat direction. These new atomic arrangements directly alter the electronic structure. Away from the interface it is characterized by carriers in the $d$-manifold with orbital symmetries $d(x^2-y^2)$ (YBCO) and $d(z^2-r^2/x^2-y^2)$ for LCMO (\autoref{fig:hetero_anatomy}, upper left), but near the interface the $d(z^2-r^2)$  become occupied in the YBCO and acquire more  $d({x^2-y^2})$ character in the LCMO.

In addition to the structural effects, an electronic mismatch occurs. The ferromagnetism in LCMO relies on the cooperative \emph{parallel} alignment of spins from the narrow correlated electronic bands; singlet Cooper pair formation in YBCO, in contrast, relies on \emph{paired} spins with antiferromagnetic interactions. These antagonistic spin  interactions (frustration) have been invoked to explain changes in the interfacial magnetization and superconductivity, $e.g.$ giant  magnetoresistance, the appearance of uncompensated magnetic moment on Cu in CuO$_2$ plane, and large modulation of   ferromagnetic magnetization profile across the heterojunction~\cite{Pena:2005p42538,Stahn:2005p42539,Chakhalian:2006p42181,Hoppler:2009p24795}

The different valence configurations of the cations in the constituent materials of the heterostructure also  induce changes in charge density and chemical bonding.  In the system shown in \autoref{fig:hetero_anatomy} (lower right panel) a charge of $\sim$0.2$e$ per Cu ion is transferred from Mn to Cu ions across the interface \cite{Chakhalian:2007p25272}.
The charge transfer at other oxide interfaces has also been found to  exhibit  a peculiar  asymmetric electronic ``roughness'' intertwined with an asymmetric interface stacking sequence or an asymmetric chemical
roughness \cite{Hoffmann:2005p41603,May:2008p7133,Chien:2013p42266}.
The effects from different stacking sequences and electronic roughness remain to be resolved.
To summarize, the following degrees of freedom are highly tunable at an oxide interface and may be exploited in uncovering new phases:
\begin{itemize} \setlength{\itemsep}{0pt}
\item Epitaxial strain mismatch owing to differences in equilibrium lattice parameters
\item Atomic coordination frustration and cation site preferences
\item Ordered spin and orbital states
\item Charge flow across the interface (layer dipole discontinuities)
\item Chemical frustration and interlayer mixing
\end{itemize}
The following examples detail how these considerations are made,
and the exciting new phases born from the interplay of the
correlated electronic and atomic structure across oxide interfaces.
\begin{figure*}
\centering
\includegraphics[width=1.98\columnwidth]{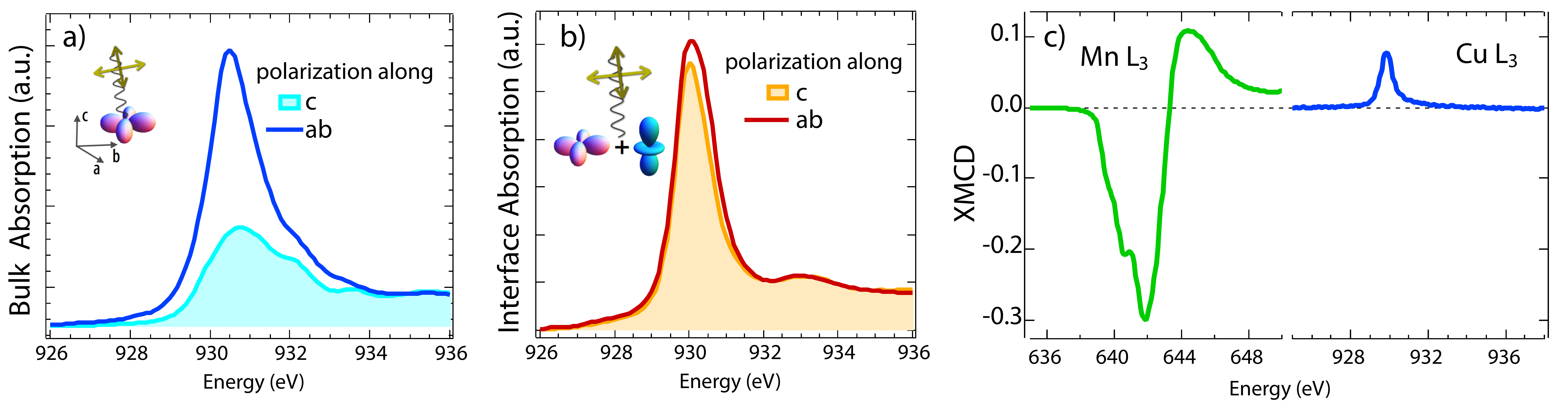}\vspace{-0.5\baselineskip}
\caption{Electronic structure of Cu in a La$_{2/3}$Ca$_{1/3}$MnO$_3$/YBa$_2$Cu$_3$O$_{7-x}$ heterostructure, determined from X-ray Linear  Dichroism (XLD) and X-ray Magnetic Circular Dichroism (XMCD) measurements. Panels (a,b): XLD spectra  taken on the Cu $L_3$-edge at temperature $T=15$ K with the electric-field vector E$\|ab$ plane and E$\|c$ plane, taken in  bulk (panel a)  and  interface (panel b) sensitive modes. The main peak (``white line'') in (b) is  shifted towards higher energies, indicating a lower charge state of Cu at the interface. Panel (c): XMCD spectra measured at the Cu and Mn $L_3$ edges in (c) recorded at $T=15$ K in a 5 T applied magnetic field demonstrating that the interfacial copper cations exhibit a non-zero ferromagnetic local moment, whereas in bulk the antiferromagnetic coupling leads to a net magnetization of zero.
\label{fig:lsmo_xas}
}
\end{figure*}
\section{Charge at the interface}
%
Understanding  and controlling  the distribution of charge carriers at  the interface between dissimilar semiconductors is one of the pivotal developments of modern microelectronics \cite{Gertner:2013} important both for devices and as a  crucial platform for  discovery of remarkable physical  phenomena including integer  and fractionally quantized  Hall effects as well as spin-Hall and other spintronic phenomena.

In conventional semiconductor heterojunctions the basic physics is driven by the difference in work-function, which causes charge transfer across the  boundary to  equalize chemical potentials. The work-function difference may be manipulated by a process known as  $\delta$-doping~\cite{Schubert:1990p42508,Harris:1991p42509}, in which a layer of ions is implanted in a plane at some distance from the interface. An additional advantage of $\delta$-doping is that the placement of the dopants at some distance from the interface minimizes the effects of randomness in the dopant positions.   $\delta$-doping is now widely  used to produce two dimensional electron gases  (2DEGs)  confined to the  proximity to the interface ($e.g.$ GaAs/AlGaAs).
The  interest in using TMO to explore similar physics was  motivated by two observations \cite{Ahn:2003p42529,Ahn:2006p42528}: ($i$) in oxides, the accessible carrier density is expected to  be orders of magnitude higher  than that of semiconductors ($\geq 10^{20}$ cm$^{-3}$), and ($ii$)  the Thomas-Fermi screening length is expected to be much shorter, so the charges may be confined to within $\textless$1-2 nm of the interface, a factor of $5-10$ shorter than the $\sim$10 nm length characteristic of  semiconductor junctions. However, the current intense effort  in material synthesis, theory, and device fabrication of oxide interfaces is motivated mainly by the known sensitivity of the correlated electron properties of transition metal oxides to the $d$-band filling ~\cite{Tokura:1999p42470,Dagotto:2001p42472,Ovchinnikov:2003p42474,Mackenzie:2003p42475,Basov:2005p42476,Lee:2006p42477,Tokura:2006p42471,Armitage:2010p40289}. The discovery of an interface-based method of carrier doping has revived the idea of tailoring the materials electronic properties and creating novel quantum states not easily attainable in the bulk counterparts. The basic idea (analogous to that motivating $\delta$-doping) is  to  explore   electronic and magnetic  phases without the hindering effects of chemical disorder inherent in the conventional solid state chemistry methods of  changing carrier concentration.

During the   past  several  years, extensive experimentation has established that perovskite-based  heterostructures are particularly susceptible to interlayer charge redistribution derived from the incompatibilities illustrated in \autoref{fig:hetero_anatomy} making them ideal candidates to explore such possibilities ~\cite{Bibes:2011p39683,May:2009p42138,Okamoto:2004p42312,Ohtomo:2004p42478,Ohtomo:2002p42318}.

\subsection{Interface Doping of a High-$T_c$ Superconductor}
To illustrate the inherent interest of charge reconstruction on  interfacial states, we discuss as one of many possible examples the recent  progress  on cuprate/manganite heterointerfaces. Macroscopically it has been established that  the introduction of a ferromagnetic (La,Ca)MnO$_{3}$ manganite layer into the heterostructure with an optimally  doped YBCO cuprate triggers a  \emph{suppression} of the superconducting transition temperature accompanied by a reduced ferromagnetic Curie temperature ~\cite{Satapathy:2012p39966,Driza:2012p41609,Kalcheim:2011p34692,Hoppler:2009p24795,Pena:2004p41605,Holden:2004p40634,Sefrioui:2003p42479}. In a recent set of experiments (\autoref{fig:lsmo_xas}),  $L$-edge polarized resonant X-ray absorption spectra taken at the Mn and Cu edges reveal the  presence of a chemical shift implying a flow of electronic charge across the interface of about $\sim$0.2\,$e$ per Cu atom \cite{Chakhalian:2007p25272,Chien:2013p42266}.  The depleted electrons from MnO$_{2}$ layer are directly transferred to  the CuO$_{2}$ planes, unbalancing the charge distribution between the atomic CuO$_{2}$ layers and the CuO chain charge reservoir block. The average Mn valence also increases from the as-grown value (Mn$^{+3.33}$) to around 3.5, indicative of covalent bond formation across the Mn--O--Cu interface.
The charge transfer across the interface from the Mn to Cu ions induces a major reconstruction of the $d$-orbital occupancies and frontier orbital symmetries in the interfacial CuO$_{2}$ layers \cite{Chakhalian:2007p25272,Chakhalian:2006p42181}.
In particular, the Cu $d_{3z^2-r^2}$ orbital, which is fully occupied and electronically
inactive in the bulk cuprates becomes active at the interface (\autoref{fig:lsmo_xas}b). At the same time charge transfer is observed in the presence of enhanced covalent chemical bonding across the interface,  the Cu cations from the nominally  antiferromagnetic CuO$_{2}$ plane acquire an uncompensated magnetic moment (\autoref{fig:lsmo_xas}c), attributed to spin canting of the local moments on the interfacial Cu cations.

\begin{figure}[b]
\centering
\includegraphics[width=0.98\columnwidth]{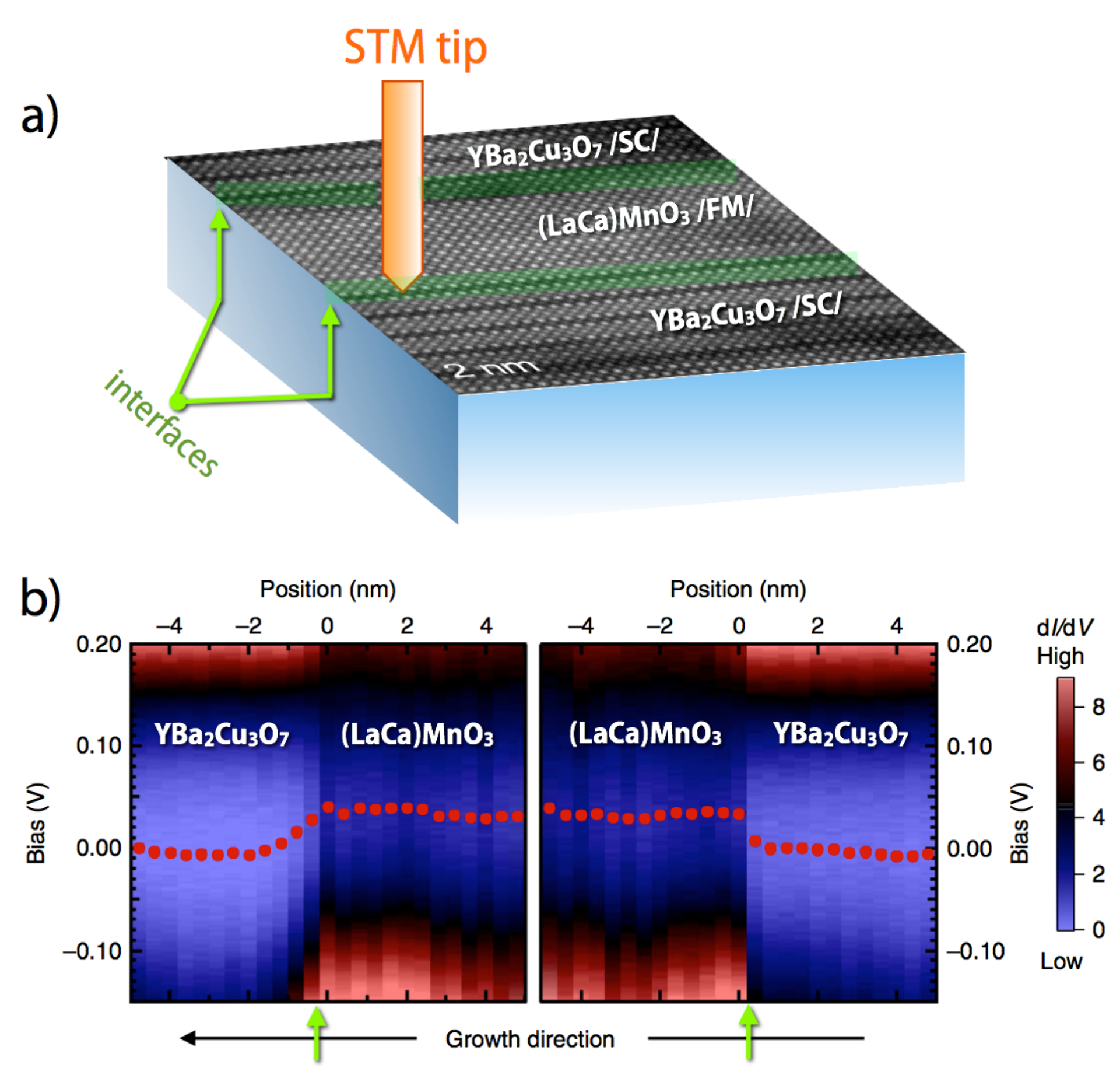}\vspace{-0.5\baselineskip}
\caption{Panel (a): Schematic of cross-sectional scanning tunneling microscopy (XSTM)
measurements performed on an LCMO/YBCO superlattice grown on a Nb-doped STO substrate.  Panel (b): Data reproduced from Ref.~\onlinecite{Chien:2013p42266}. The spatial evolution of the $dI$/$dV$ spectra averaged across the two identically terminated heterointerfaces reveals that the electronic transition is more abrupt for the bottom interface (right arrow) than the top, broader, interface (left arrow).
The red dots represent the voltage of the minimum in the density of states.}
\label{fig:XSTM}
\end{figure}

Initial studies of the interplay between the ferromagnetic and superconducting order parameters used synchrotron based  X-ray and neutron reflectivity experiments. However these tools were unable to clearly resolve the length scale of interactions at the boundary between the two phases. Very recently, the issue has been addressed by use of  cross-sectional scanning tunneling microscopy (XSTM)  together with atomic-resolution electron microscopy (EELS/STEM). These methods enable direct observation of the charge distribution and the corresponding spatial  scale for the buried interface \cite{Chien:2013p42266}. \autoref{fig:XSTM} shows the spatially resolved $dI$/$dV$ spectra, which provided the first direct evidence that the length scale for charge transfer between YBCO and LCMO has an upper limit of \textless 1 nm,  and that the spatial broadening of the electronic transition is commensurate with the rougher interface.
This result sets a fundamental upper limit on the charge-transfer length scale in the YBCO/LCMO system, ruling out a class of theories based on long-range proximity effects~\cite{Hoffmann:2005p41603}. In addition to the established X-ray  and neutron based probes, this   powerful characterization technique provides a useful tool to  achieve a microscopic direct space understanding of the electronic structure across correlated oxide interfaces.

\subsection{Additional considerations}
The complex behavior occurring at the LCMO/YBCO interface highlights the need to develop a clear language and set of concepts to describe interface electronic physics in correlated oxides. The inherently many-body nature of the correlated interface raises
fundamental questions,  in particular of the applicability of the ideas, formulae,  and language devised for semiconductor interfaces where a single-particle description works well. Pioneering work of Oka and Nagaosa \cite{Oka:2005p42264} showed via density matrix renormalization group calculations of a one dimensional model system (in essence the one dimensional Hubbard model with a spatially varying interaction parameter and band bottom) that  the standard concepts of band bending and interface dipole apply, albeit with some modifications, as long as the conduction and valence bands are replaced by lower and upper Hubbard bands.
A growing body of literature builds on this work, using the concepts of band bending, Schottky  barriers, and depletion layer creation  borrowed from semiconductor physics \cite{Yajima:2011p27821,Hikita:2009p11794},  as well as more involved  approaches, which unite Poisson-Schr\"odinger electrostatics with Mott-Hubbard physics
\cite{Okamoto:2004p42312,Charlebois:2013p42487,Lee:2006p42486}.  Correlation physics is shown to lead to \emph{quantitative} changes in the spatial confinement of carriers near interfaces \cite{Okamoto:2004p42312,Lee:2006p42486,Lee:2007p42491}, including the possible formation of extended depletion regions of zero  compressibility  (so-called `Mott plateaus')\cite{Lee:2006p42486,Charlebois:2013p42487}. Other theoretically proposed possibilities
\textit{unattainable with semiconductor junctions}, include a spontaneously emerging quantum-well structure when an electron-doped Mott-Hubbard insulator is coupled to a normal metal  with  a large work-function. Following the same line of reasoning, in a $p$-$n$ junction between two correlated insulators the local Mott gap collapses giving rise to a 2DEG~\cite{Charlebois:2013p42487}.

With few exceptions \cite{Jin:2011p42490}, current experimental attention has focussed  on interfaces such as that between the two band insulators LaAlO$_3$ and SrTiO$_3$. In most of these situations the carriers are introduced via the polar catastrophe mechanism \cite{Mannhart08}; the maximum sheet carrier density is $0.5$ per in-plane unit cell and this carrier density is typically distributed \cite{Mannhart08,Okamoto:2006p42492} over several unit cells away from the interface, leading in general  to volume carrier densities far below the Mott value of one per unit cell. Density functional plus Hubbard $U$ calculations \cite{Pentcheva07}  indicate that a charge ordered phase in which the entire polar catastrophe charge density is in the first interface layer may be possible, but these suggestions have not yet been confirmed by experiment or beyond-DFT methods. One very interesting potential exception is the work of \onlinecite{Moetakef12} on GdTiO$_3$/SrTiO$_3$ heterostructures, where a nontrivial insulating phase was observed when two layers of SrTiO$_3$ were sandwiched between thick sheets of GdTiO$_3$. It has been explained by Chen, Lee and Balents in terms of a novel `Mott dimer' phase \cite{Chen13}, where the carrier density is far below the one electron per transition metal ion value needed for Mott physics; nonetheless many theoretical predictions suggest alternative avenues for emergent properties to arise and  warrant experimental investigation.
Additional issues beyond conceptual approaches to interface control arise. The length scales in correlated oxides are typically very short, so the details of the interface may be more important than in conventional semiconductors.A local picture is needed, which is able to address the formation of chemical bonds across the junction, differing electronegativities of transition metal ions, changes in both crystal field energies and Madelung  potentials, and polarity effects \cite{Salluzzo:2013p41953,GarciaBarriocanal:2013p42121,Park:2013p41899,Biscaras:2012p42167,Zhong:2010p42484,Savoia:2009p42485,Takizawa:2009p4135,Sing:2009p2694,Herranz:2007p42483,Hotta:2007p42258,Ohtomo:2004p42478}.

A further complication is that while many correlated oxides are reasonably well described by the Mott-Hubbard picture on which the above-cited works are based, some important functional TMO are \emph{charge-transfer} compounds ~\cite{Imada:1998p38738,Khomskii:1997p42493,Zaanen:1985p4144}. The role of the lower Hubbard band in these materials is usurped by the ligand states (typically oxygen $2p$), thus implying a very different physical character for the doped holes (mainly in oxygen levels) and doped electrons (mainly in transition metal $d$-levels). As a result, the alignment of the oxygen levels across the interface becomes crucial.

For all of the materials discussed in this paper, theoretical treatments which go beyond the simple Hubbard model, including chemically realistic structures and energetics on the same footing as correlation effects, are needed, as are experimental investigations of systems with higher electron densities and complete control over cation and oxygen stoichiometry.

\section{Control of Magnetism with Oxide Heterostructures}
Long range magnetic order in transition metal oxides usually arises from a combination of  local moment formation on the transition metal site and inter-site coupling via the oxygen sublattice.  Heterostructures offer an opportunity to generate new magnetic states by manipulating both the moment formation and the nature of the inter-site coupling.  As examples, we note that the paramagnet LaCoO$_3$ can be converted to a ferromagnetic (FM) material by tensile epitaxial strain, which changes the material from a low-spin to a high-spin state   \cite{Freeland:2008p25261,Park:2009p25266,Rondinelli:2009p1318,Fuchs:2007p10350}.
On the other hand, bulk antiferromagnetic (AFM) EuTiO$_3$ can be converted to a ferromagnetic  insulator under modest tensile strains \cite{Lee:2010p25085}. Another notable example is the comprehensive study by  \onlinecite{Seo:2010p42273}, which examined three-component SrRuO$_3$/manganite/SrRuO$_3$ heterostructures. These authors found strong compressive strain causes relative FM alignment of magnetization in the heterostructure layers, while tensile or weak compressive strain favors AFM alignment of neighboring layers.

This sort of  control over local magnetization in thin film geometries is of potential utility for oxide electronics and spintronic applications, including magnetic memory and sensing \cite{Bibes:2011p39683}. For example, electromechanical coupling via a piezoelectric
material can be used to control the orientation and strength of the magnetization by tuning the lattice parameters of the heterostructure through an applied electric field \cite{Dekker:2011p34081}.  Here, we focus on going beyond strain control to make use of the
broken symmetry at the interface between two dissimilar materials to generate unique spatially structured magnetic states.
\subsection{Creating Novel Magnetic States at Interfaces}
\begin{figure}
\centering
\includegraphics[width=0.98\columnwidth]{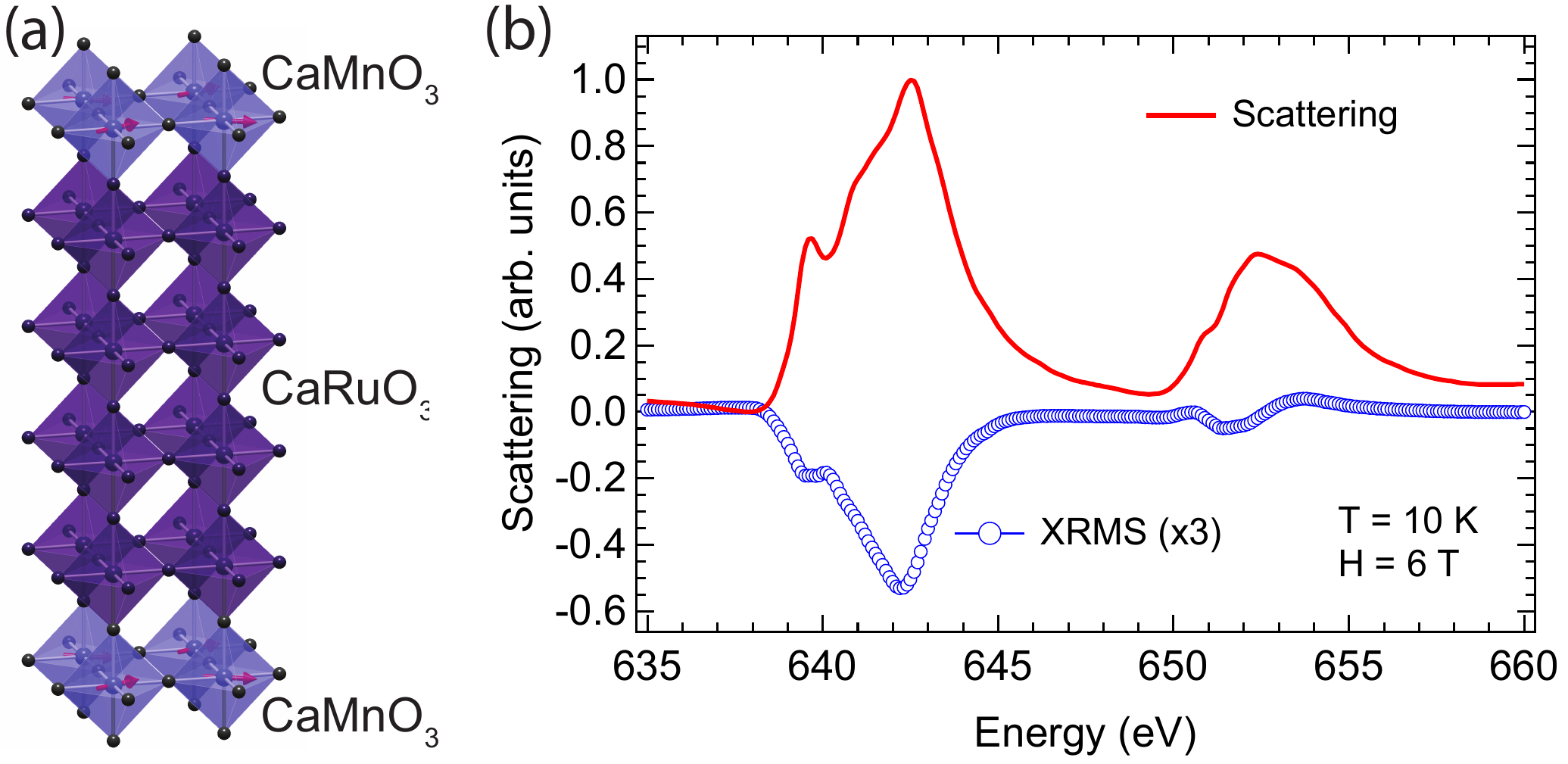}\vspace{-0.2\baselineskip}
\caption{
Panel (a): Schematic crystal structure showing  canted spins (pink arrows) within the MnO$_6$ octahedra of CaMnO$_3$ at the interface of the quantum-well structure with metallic CaRuO$_3$. The canting arises from electron transfer owing to the ohmic contact. Panel (b): X-ray resonant magnetic scattering data showing a large magnetic signature arising from the FM alignment of spins at the interface in the presence of a magnetic field (see associated data in Ref.~\onlinecite{Freeland:2010p25254}).
\label{fig:CROCMO}
}
\end{figure}

One approach to manipulating magnetism involves interfacial charge transfer in  heterostructures created from an antiferromagnetic \emph{insulator} and a paramagnetic \emph{metal} \cite{Takahashi:2001p42254,Freeland:2010p25254,Yordanov:2011p32557}. The choice of materials in this case was determined by two key factors: first, creating moments from a material without any propensity to moment formation, $i.e.$, zero moments, is difficult. It is therefore reasonable to  begin then by choosing  a system with a large local moment such as CaMnO$_3$ with $3\mu_B$/Mn, which in bulk is a G-type (conventional two-sublattice N\'eel) antiferromagnet. However, modest electron doping of this material leads to strong ferromagnetic (FM) correlations \cite{Neumeier:2000p42400}.  In a quantum-well heterostructure in which a paramagnetic metal (in this case CaRuO$_3$) is confined between two thick layers of CaMnO$_3$ one may expect that charge transfer from the metal to insulating CaMnO$_3$ will lead to interfacial doping and thus ferromagnetism.
Theoretical studies substantiate this argument and find that  a charge of approximately $0.1$\,e per interface unit cell leaks across the interface and is  confined within $\sim$1 unit cell at the CaRuO$_3$/CaMnO$_3$ interface \cite{Nanda:2007p12003}. Although the magnitude of the charge leakage is small, it has a significant impact on the antiferromagnetic order in the CaMnO$_3$, providing a mechanism for spin canting which yields large ferromagnetic moments at the interface \cite{Takahashi:2001p42254,Freeland:2010p25254,Yordanov:2011p32557}. To validate this concept a study of the spatial distribution of the magnetism was carried out using X-ray resonant magnetic scattering (XRMS) at the Mn $L$-edge \cite{Freeland:2005p25285,Kavich:2007p25275}. \autoref{fig:CROCMO} shows the large XRMS signal, and that it deviates from anticipated bulk G-type AFM state, which shows no ferromagnetic component to the magnetic moment under identical strain conditions indicating that the ferromagnetism emerges from the interface \cite{Freeland:2010p25254}. By fitting this signal as a function of incident angle, the extent of the magnetic polarization away from the interface was found to extend over several unit cells in contrast to the length of one unit cell predicted by theory \cite{Nanda:2007p12003}.  The observed longer length scale of the magnetization profile discrepancy may be due to magnetic polarons, which are known to exist in lightly-doped CaMnO$_3$ \cite{Chiorescu:2007p42405}, but such interfacial polarons have not explicitly investigated theoretically.

\subsection{Other Routes to Interface Magnetism}
Strain and layer sequencing can  offer additional handles to manipulate the interfacial magnetic state in the CaRuO$_3$/CaMnO$_3$ system \cite{He:2012p41809}. The link between the metallic layer and magnetism is best illustrated by studying superlattices where the metallic layer undergoes a metal-to-insulator transition when the dimensionality is reduced in the ultra-thin layer, and correspondingly the magnetism disappears \cite{Grutter:2013p42329}. One can use this understanding and exploit it to design new functional materials and there are many possibilities that exist within the perovskite familiy which can be combined to seek new types of magnetic states \cite{Smadici:2007p29592,Bhattacharya:2008p7251,Gibert:2012,Hoffman:2013p42403}. For example, many antiferromagnets have ordering temperatures well-above room temperature, so one could extend this concept to create interfacial insulating \emph{ferri}magnets that operate at high temperature \cite{Ueda15051998}.
Another possibility is to make use of the spatially localized magnetic state in proximity to a metallic layer to create a spin-polarized 2DEG \cite{Nanda:2008p54}. More broadly, one could create heterostructures with two magnetic materials, and use the competition towards different collectively ordered magnetic states in addition to structural
incompatibilities to generate a plethora of interesting and potentially spatially varying magnetic phases. These are  but a few of the magnetic possibilities which remain to be
uncovered at oxide heterointerfaces, chosen to highlight the large phase space still available for exploration and the opportunities available to connect with materials theory in the rational search for new magnetic systems.
%

\section{Interfacial Control of Orbital Polarization}

\subsection{The Case of Rare-earth Nickelates}
The orbital configuration, $i.e.$, the distribution of the $d$-electrons over the available crystal field levels,  plays an important role in the formation of strongly correlated ground states in transition metal oxides \cite{Tokura:2000p42052}. In general, orbital configurations are  closely linked to structure and may therefore be manipulated at interfaces.  Here we discuss these issues specifically for the orthonickelate perovskites $R$NiO$_3$, where $R$ is a trivalent cation from the lanthanide  series, but the ideas can be extended to other $AM$O$_3$ systems.

The original and decade later renewed interest in nickelates arose from the possibility of generating a cuprate-like electronic and orbital configuration in a copper-free system
\cite{Hamada93,Anisimov:1999p30820,Lee:2004p30821,Chaloupka:2008p288,Poltavets:2010p30805}. The basic idea is that in bulk $R$NiO$_3$ the Ni is octahedrally coordinated, with only small deviations from cubic ($O_h$) symmetry. Further, formal valence considerations indicate that the nominally Ni$^{3+}$ cation is in the low-spin $d^7$ configuration, with the $t_{2g}$ states ($d_{xy,xz,yz}$) filled and one electron in the two-fold degenerate $e_g$-symmetry ($d_{3z^2-r^2,x^2-y^2}$) Ni $d$-levels. Low-spin $d^7$ is a first-order Jahn-Teller configuration, with a  susceptibility to bond distortions which break the cubic point symmetry and are enhanced by correlation effects. It was thus expected that modest perturbations would split the $e_g$ levels, leaving an effective one-band configuration where the electron is fully confined to a single orbital.

The degree to which an electron occupies two different  $m_{l1}$ and $m_{l2}$ orbitals
can be quantified as an \emph{orbital polarization}
\[
P_{l_1 m_{l1},l_2 m_{l2}}=\frac{n_{l_1 m_{l1}}-n_{l_2 m_{l2}}}{n_{l_1 m_{l1}}+n_{l_2 m_{l2}}}\, ,
\]
where $n_{l_1 m_{l1}}$ and $n_{l_2 m_{l2}}$ are the occupancies of the  $\left|l_1 m_{l1}\right\rangle$ and $\left|l_2 m_{l2}\right\rangle$ states \cite{Han:2010p28554},
with orbital quantum number $l_i$ 
and magnetic quantum number $m_{li}$,  
respectively.
For the rare-earth nickelates, the relevant orbital polarization arises from the
$n_{x^2-y^2}$ and $n_{3z^2-r^2}$ occupancies, and a fully polarized state $P=1$ would
be indicative of a single band electronic structure.

Something akin to this effect occurs in many members of  the ``colossal'' magnetoresistance manganites, where the basic configuration is a high-spin $d^4$ configuration and similarly a Jahn-Teller ion that can be manipulated with strain \cite{Tokura:2000p42052}. Hubbard-model calculations further indicated that the single-band physics was very likely to appear \cite{Hansmann:2009p12185}; however, more realistic
\emph{ab-initio} calculations indicate that the actual electronic configuration for Ni is in the high-spin $d^8$ state with a hole on the oxygen ($d^8\bar{L}$) \cite{Han:2010p28554}. Since the high-spin $d^8$ configuration has one electron in each of the two $e_g$ orbitals, it is significantly less susceptible to undergoing Jahn-Teller distortions,
suggesting that it would be more difficult than initially expected to achieve the
desired degree of orbital polarization, even in the correlated case \cite{Han:2011p41333}. Studies of the dependence of orbital polarization on the different flavors of structural symmetry-breaking \cite{PhysRevB.87.155135} is thus of great experimental interest and is a stringent test of the theory.
\subsection{Manipulating Orbitals in $R$NiO$_3$ Heterostructures}
Advances in high-quality growth of nickelates over the past few years  mean that we are now in a position to test these predictions \cite{Tsubouchi:2008p179,Eguchi:2009p2955,Scherwitzl:2010p35780,Liu:2010p25255,May:2010p18583,Boris:2011p35832,Hwang:2013p42503,Bruno:2013p42438}. The basic experimental approach is to use a combination of quantum confinement, achieved by fabricating ultra-thin layers of TMO sandwiched between layers of wide-gap insulators, and epitaxial strain, obtained by varying the substrate material, to break the octahedral symmetry. Advanced x-ray techniques are then used to estimate the resulting changes in orbital occupancies.

However, \emph{ab-initio} calculations based on density functional theory indicate that the contribution of strain to octahedral symmetry breaking is not completely intuitive (see \onlinecite{Rondinelli:2012p42180} and references therein).  In particular, a considerable degree of compression or tension can be accommodated by octahedral rotations, without necessarily changing the local point symmetry significantly since the NiO$_6$ units are
highly flexible \cite{Chakhalian:2011p34974}. Furthermore, quantum confinement may be affected by the chemistry of the insulating layer, with different degrees of polarization found for different choices of wide-gap insulator \cite{Han:2010p28554}.

At  present, the experimental results are not completely consistent with each other or with theory.  For example, examination of the Ni $L_2$ edge indicated an  $\sim5\%$ orbital polarization for a single unit-cell of LaNiO$_3$ subject to tensile strain \cite{Freeland:2011p42179} and no orbital polarization for compressive strain. Other measurements employing an orbital reflectometry technique on four unit cell films also observed a similar non-zero interfacial polarization for tensile strain \cite{Benckiser:2011p27820,Frano:2013p42355}. Recent studies have indicated it is possible to increase the orbital polarization  up to $25\%$ through judicious optimization of high tensile strain  states and alternative spacer materials \cite{Wu:2013p42419}; the latter had been shown theoretically to play a considerable role in obtaining the targeted orbital polarization levels \cite{Han:2010p28554}.
\subsection{Open Questions in Orbital Control at Interfaces}
All experiments agree though that the degree of orbital polarization observed in actual superlattices is small compared to that needed to achieve a fully orbital polarized Ni $e_g^1$ state. The main challenge is to then build the framework to understand how to create fully orbital polarized states in oxide heterostructures.

One important facet of this problem has to do with strain and symmetry. For example, LaNiO$_3$ has rhombohedral symmetry in the bulk which actually disfavors a uniaxial Jahn-Teller distortion \cite{Carpenter:2009p3987}. NdNiO$_3$, on the other hand, is orthorhombic which allows such a distortion without large energetic penalties. Recent studies by \onlinecite{Tung:2013p42442} show that the nickelate films maintain to some extent the symmetry of the bulk, which, due to the connection between compatible lattice distortions and crystal symmetry, directly influences the ability to orbitally polarize the $3d$-states even under large strains.

With this understanding, one may be able to choose the proper bulk symmetry of the TMO to be used in the heterostructure to build in larger orbital polarizations in NdNiO$_3$ by coupling strain with the interfacial covalency effect discussed above and interfacial proximity effects \cite{doi:10.1021/nl500235f,doi:10.1021/cg500285m}. Even for the case of NdNiO$_3$ films, however, the orbital polarization is still insufficient to create a fully polarized state \cite{Tung:2013p42442}. This is largely due to the energy scale mismatch between elastic strain ($\sim$ 100 meV) and the bandwidth (on order of several eV),  and the overall tendency to orbital polarization is further reduced by the $d^8\bar{L}$ character of the Ni$^{3+}$ state.
Small orbital polarizations have also been observed even in the case of the Jahn-Teller active manganites \cite{Aruta:2006p42142,Tebano:2008p42139,Pesquera:2012p42127}, which indicates that this balancing of drastically different energy scales is difficult even in systems that prefer orbital order.  A potential solution is to create interfaces with large symmetry mismatch due to lattice topology or by combination of dissimilar crystal field environments.

Consider for example  bulk oxides with large orbital polarization such as the cuprates \cite{Nucker:1995p42064,Chen:1992p42068} and Ruddlesden-Popper (layered-structure)  nickelates \cite{Kuiper:1998p42189,Pellegrin:1996p42182} as a starting point.  In these materials, the large orbital polarization arises from the strongly asymmetric crystal (ligand) field of the layered structure. As was discussed above for the LCMO/YBCO heterointerface, oxide interfaces can be harnessed to `undo' orbital polarization, but there is no reason why the converse should not also be possible. This offers a real opportunity in the area of matching systems with  drastically different symmetries to create orbital states at the interface.
Orbital control can also be used to modulate strongly correlated states. Strain very effectively controls the metal-insulator transition (MIT) for NdNiO$_3$ thin films \cite{Liu:2010p25257, Liu:2013p42440}, but the underlying mechanism is not fully understood. Using quantum confinement when the layer dimensions approach the atomic limit, it was observed that orbital polarization under compressive strain tends to favor a metallic state while quantum confinement caused a re-emergence of a MIT through the interfacial reduction of the orbital polarization \cite{Liu:2012p41668}.
A similar connection was recently observed in the case of VO$_2$ thin films \cite{Aetukuri:2013p42392}, where the decrease in the MIT temperature was correlated with strain driven polarization of the V $t_{2g}$ orbitals. The potential use of strain in combination with symmetry mismatch to tune between correlated metallic and insulating phases is an important issue warranting further investigation.
 \section{Ferroelectric Heterostructures from Nonferroelectric Bulk Oxides}
The electrically switchable polarization of ferroelectrics (FE) allows their integration in random access memories (FE-RAM), electro-optical devices, sensing microsystems, active vibration control and surface acoustic wave systems, to high frequency devices~\cite{Setter:2006p42422}.  The main challenges for future FE-RAM scaling, however, is that the FE dielectric thickness must be reduced to fit within the required device area while maintaining sufficient reproducibility and signal margins for sense amplifier differentiation
between a `0' and `1' data state~\cite{Wu:2010p42427}. Furthermore, non-destructive \emph{magnetic} sensing of \emph{electric} polarization, enhanced miniaturization and increased packaging density in magnetoelectric materials (ME)
\cite{Fiebig:2005p42423,Eerenstein:2006p42271,Ramesh:2007p42429,Velev:2011p42270} would enable the realization of four-state logic in a single device \cite{Bibes:2008p12103,Khomskii:2009p42424}.

\begin{figure}
\centering
\includegraphics[width=1.\columnwidth]{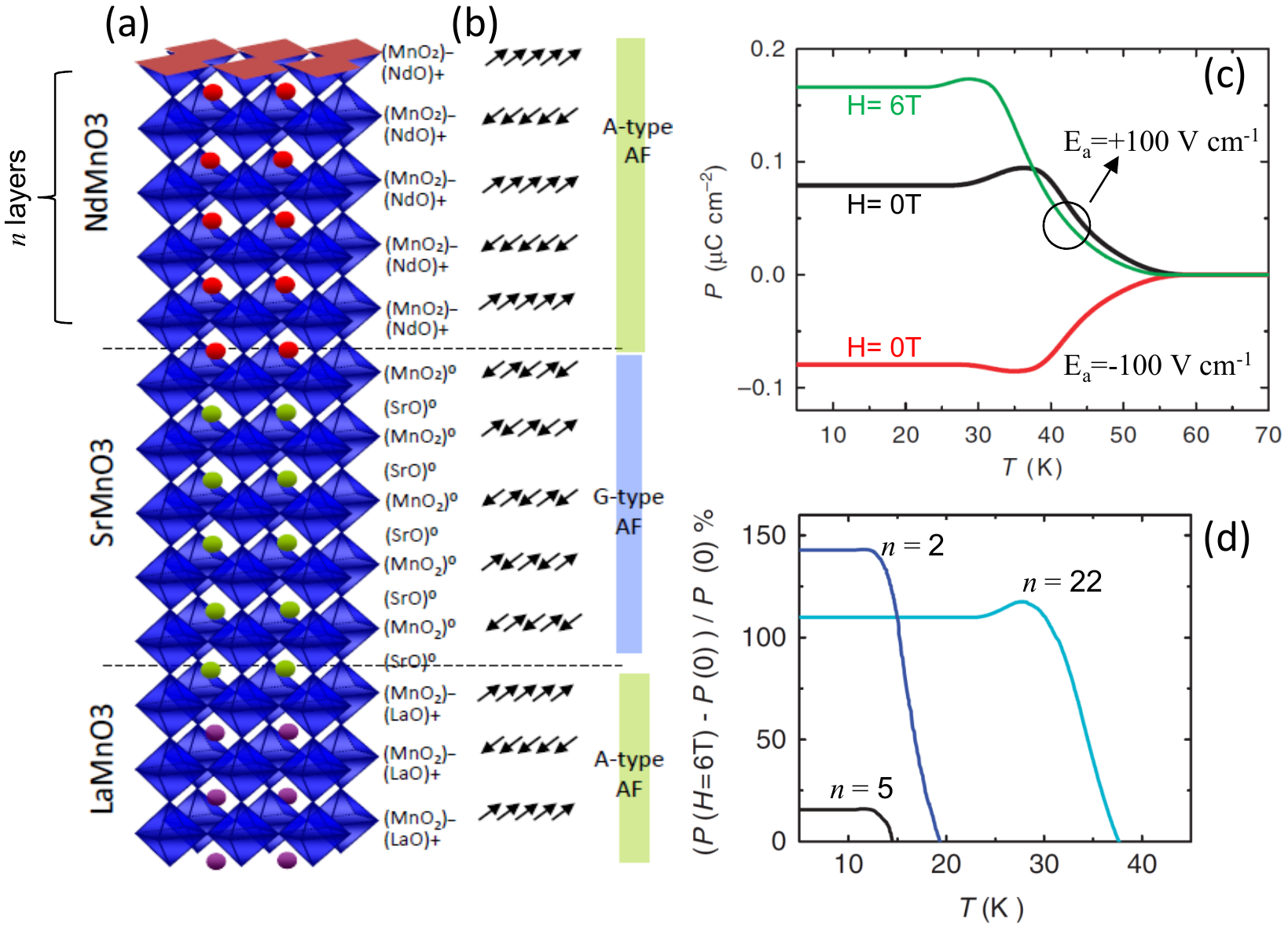}
\caption{%
Ferroelectric and magnetoelectric properties of  [(NdMnO$_3$)$_n$/(SrMnO$_3$)$_n$/(LaMnO$_3$)$_n$]$_m$ superlattice, where (n,m) denotes the specfic superlattice structure (a) Schematic
[(NdMnO$_3$)$_5$/(SrMnO$_3$)$_5$/(LaMnO$_3$)$_5$]$_8$ superlattice on single-crystalline SrTiO$_3$ substrate with the metal--oxygen octahedra and A cations emphasized.  The arrays of the arrows in (b) represent the corresponding antiferromagnetic spin arrangements for each component of the heterostructure. (c) Temperature ($T$) dependence of the electric polarization ($P$) measured in a superlattice of period (22,2)   using the pyroelectric technique for a typical electric field (E$_\mathrm{a}$) of +100 V cm$^{-1}$ (black curve) and $-$100 V cm$^{-1}$ (red curve) applied
perpendicular to the plane of the superlattice layering.  The temperature-dependent electric polarization under a magnetic field $H$=6~T applied parallel to the plane of the
superlattice layering (green curve) reveals strong magnetoelectric coupling.  (d) Normalized relative change in the electric polarization at fixed electric and magnetic fields for various superlattices. Figure adapted from Ref.~\onlinecite{Rogdakis:2012p42268}.  }
   \label{fig:FE}
\end{figure}

The conventional approach for realizing strong ME materials, i.e., where their is strong coupling between the primary electric and magnetic polarizations, uses naturally occurring materials possessing primary ferroic orders, namely ferroelectricity and ferromagnetism.  Such materials not only are rare, but often suffer from weak coupling between the spin and charge degrees of freedom \cite{Eerenstein:2006p42271}.
Recent advances in atomic layer epitaxy now enable the design and fabrication of heterostructures with atomically flat interfaces that can support new forms of ferroelectricity \cite{Bousquet:2008p42309,Mulder/Benedek/Rondinelli/Fennie:2013,Rondinelli/Fennie:2012} and magnetoelectric coupling owing to interfacial interactions among electronic spins, charges, and orbitals \cite{Wu:2010p42427}.  A promising avenue to pursue in the search for new materials with emergent ferroelectricity and a strong magnetic field dependence of the electric polarization exploits an superlattice structure with broken inversion symmetry, which results from being constructed from three distinct layers~\cite{Warusawithana:2003p42526,Lee:2005p42527}. The `tri-color' layering lifts inversion symmetry -- a prerequisite for an electric polarization -- whereas epitaxial strain applied to the heterostructure can promote the formation of electrically and magnetically tunable polarizations, even in the absence of ferroic components~\cite{Tokura:2007p42426,Hatt:2007p42425}.
Using a combination of complementary experimental probes, magnetoelectricity was  recently demonstrated in artificial tri-layer heterostructures consisting solely of  dielectric antiferromagnetic oxides (\autoref{fig:FE}a). Laser molecular-beam epitaxy was used to create the heterostructure comprising alternating LaMnO$_3$, SrMnO$_3$, NdMnO$_3$ layers on a SrTiO$_3$ substrate. \onlinecite{Rogdakis:2012p42268} report the emergence of ferroelectricity below 40K (\autoref{fig:FE}c) and it was found to depend on the number of NdMnO$_3$ layers $n$ in the superlattice (\autoref{fig:FE}d).
Interestingly, the authors observed slim loop-like polarization--electric ($P$-$E$) field hysteresis, with an extended tail of the polarization above the ferroelectric transition temperature and a thermal hysteresis between zero-field-cooled and field-cooled measurements.
Such features are typical of relaxor ferroelectrics and were attributed  to interface effects \cite{Rogdakis:2012p42268}.
We note that this dielectric relaxation also leads to differences in the  magnitudes of the measured polarization obtained from
the $P$-$E$ loop and the pyrocurrent measurement, which might also be affected from the challenges in
characterizing the dielectric properties of ultrathin film oxides with techniques commonly used for bulk single crystals.
Nonetheless, the magnetoelectric coupling resulted in 150\% magnetic modulation of the electric polarization, demonstrating how heterostructuring multiple compounds together to lift inversion symmetry in superlattices is an avenue to create new functionalities.

First-principles density functional calculations indicated that broken space inversion symmetry and mixed valency, arising from the heterostructure geometry (cation layer sequence) and interfacial polar discontinuity, respectively, is responsible for the observed behavior.  In particular, the formal charge layering of the LaMnO$_3$ and NdMnO$_3$ components at the interfaces with SrMnO$_3$ give rise to a charge discontinuity, leading to electron transfer and cooperative off-centering of the cations.  The $A$ cation layering leads to a pattern of Mn and $A$-cation displacements along the superlattice normal growth direction that lift inversion symmetry and therefore produce the macroscopic electric polarization.
We note that the ferroelectric relaxor behavior could not be seen from the theoretical results, which capture
the static and cation ordered zero-temperature behavior.

This work demonstrates yet another fascinating example of emergent functionality exhibited in heterostructures. The ability to lift inversion symmetry and independently tune spin order allows the design of many more materials with multifunctional behavior \cite{Puggioni/Rondinelli:2014,ADMI:ADMI201400042}.  One may exploit these systems to engineer devices from artificial low-dimensional materials exhibiting novel tunable functions distinct from that of bulk systems.

\section{Conclusion}

The physics of interfaces between materials exhibiting correlated electronic behaviors including superconductivity, magnetism and ferroelectricity  is  a rapidly advancing field, situated at the intersection of materials science, solid state chemistry and condensed matter physics. Understanding and exploiting these remarkable systems places extraordinary demands on synthesis, measurement and theory, and the challenge is stimulating remarkable work in all areas.  By way of conclusion we highlight challenges and prospects in correlated oxide interfaces.

\subsection{Chemical and structural order}
Characterization and control of chemical and structural order is a crucial issue. While research to date has revealed remarkable phenomena, clearly related to properties of theoretically ideal interfaces, effects of disorder are not negligible. The brutally short length scales (often only one or two unit cells) pose strong constraints on materials quality. For example, metal to insulator transitions generically occur in oxide heterstructures when the thickness of the metallic layer becomes of the order of 1-2 unit cells. Systematic dependence on strain \cite{Son:2010p14460}, and systematic evolution of electronic structure with thickness \cite{Yoshimatsu:2011p32626} suggest an important intrinsic component, but disorder effects and changes in growth processes on these length scales cannot yet be ruled out as mechanisms.  Antisite defects mean that real interfaces are not as sharp as depicted in the idealized sketches shown in this paper, and these defects are not necessarily easy to identify in transmission electron microscopy experiments, which average over columns of order $10^3$ atoms. Further, oxygen defects and interstitials play a crucial role in transition metal oxides and oxygen partial pressure during growth and in post-growth annealing of heterostructures clearly affects properties in many cases  \cite{Nakagawa:2006p12453,Ariando11}. Methods to further define and control the actual structure of interfaces are urgently needed. One area of future study, is to couple the insight from {\it in-situ} studies of oxide film synthesis to that of multiscale theory in order to build a mechanistic understanding of the process by which interfaces are created.

\subsection{Theory}
The importance and interest of oxide interfaces for the general issue  of the theory of correlated electron materials cannot be overemphasized. Understanding the phenomena at interfaces requires a combination of sophisticated many-body physics (to understand the correlated electron states) and \emph{ab-initio} insights (to understand the implications of the changes in octahedral rotations, atomic coordination, and lattice relaxations). The present state of the theoretical art is a combination of analysis of model systems (in particular the Hubbard model), which cannot easily encode many real materials aspects, in particular the transition-metal/ligand covalence as well as the energetics associated with lattice relaxations and \emph{ab-initio} techniques (especially the DFT+$U$ method) which have provided crucial insights but are based on a greatly oversimplified Hartree approximation to the many-body physics and may overemphasize order \cite{PhysRevB.80.235114,PhysRevB.86.195136}.  In particular the status of the DFT+$U$ predictions of magnetism \cite{Okamoto:2006p42492} and charge order \cite{Pentcheva07} at the LAO/STO interface remains unclear.

The combination of density functional band theory and dynamical mean field theory (DFT+DMFT) is a promising alternative \cite{Kotliar06}, combining \emph{ab-initio} and many-body physics in a systematic way. However, working implementations of total energy calculations are only now beginning to appear \cite{Park13} and forces cannot yet be computed so structural optimization remains a challenge. More fundamentally, existing implementations for systems in which more than one $d$-orbital is important are based on the single-site approximation, which is believed to become poor in the two dimensional situation relevant to heterostructures.

\subsection{Topological states of matter}
Topological insulators (TIs) are a fascinating class of materials in which strong spin-orbit interaction promotes  gapless electronic states on the surface ($i.e.$ edge states)  with the bulk of a material remaining gapped~\cite{Fu:2007p42507,Hsieh:2008p42510,Moore:2010p42512,Qi:2010p42513,Hasan:2010p42511}. Most of the current TI materials belong to the Bi$_2X_3$ ($X$=Se, Te) family. Recently,  a new approach has been  proposed that is based on superlattices of two (or three) unit cells of a strongly correlated electron perovskite $AB$O$_{3}$  grown \emph{along the [111]  direction} combined with a band insulator spacer layer; the resulting heterostructure  structurally forms a buckled honeycomb lattice topologically equivalent to that of graphene lattice for the case of three unit cell strongly correlated oxide.  Depending  on the strength of electron-electron correlations, magnitude of Hund's coupling and inter-site hopping, the proposed heterostructures display potentially rich physics associated with exotic electronic and topological phases~\cite{Ruegg:2011p38511,Yang:2011p38510,Xiao:2011p41911,Ruegg:2012p41179,PhysRevLett.110.066403,Ruegg:2013p42515}.
At  present, the main challenge in experimental realization is the film growth along the [111] direction since for the commonly used substrates, $e.g.$ SrTiO$_3$ LaAlO$_3$, NdGaO$_3$, YAlO$_3$, etc., the (111) structure consists of alternating $\pm4e$ or $\pm3e$ charged planes along this direction. The large polar discontinuity generally results in complex surface/interface and electronic reconstructions  \cite{Enterkin:2010,Marks20092179}, which  can act to compensate for the polar mismatch. To date there is  limited understanding of  thin film nucleation, growth and charge compensation in perovskites along 
highly polar directions. Very recently the synthesis work in this  direction has been initiated ~\cite{middey:261602}.

\subsection{Oxygen Defect control}
While many of the examples discussed above involve  oxygen stoichiometric  perovskites, the ease of removal/addition of oxygen can also offer opportunities for materials that can be programmed by their chemical environment \cite{Kalinin:2012p41881,Kalinin:2013p42533}. While the role of oxygen vacancies has been explored deeply in the context of catalysis and fuel cells \cite{Adler:2004p41756}, recent work has highlighted the controlled stabilization of related oxygen deficient phases using oxide heterostructures. This is interesting for epitaxial thin film phases such as SrCoO$_{3-\delta}$ \cite{Jeen:2013p42437,Jeen:2013p42436} or La$_{1-x}$Sr$_x$FeO$_{3-\delta}$ \cite{Xie:2013p42532}, which can be reversibly converted between oxygen deficient and stoichiometric phases at low temperatures. Since these phases have drastically different ground states, it offers an interesting path for control of strongly correlated electrons via dynamic anion compositional control. By combining low conversion energy with electrochemical gating of vacancies, such as that seen recently for VO$_2$ \cite{Jeong:2013p42020} and $R$NiO$_3$ \cite{Shi:2013p42435}, this approach allows direct control of metal vs. insulating phase as well as possible elements of brain-like (neuromorphic) electronic circuits.

\subsection{Moving beyond the static realm}
Up to now, all the properties that have been discussed were limited to the quasi-equilibrium properties, but in the future one should also investigate the \emph{dynamical}  degree of freedom to explore the emergence of unique transient states. While the dynamic response for bulk  materials  has been extensively investigated~\cite{Averitt:2002p41326,Basov:2005p42476}, oxide heterostructures offer new possibilities. Recent pump-probe studies of oxide films illustrate the potential for ultrafast strain modulation~\cite{Daranciang:2012p41904,Wen:2013p41915}, which allows one to manipulate the lattice in a new direction since the film motion is clamped in-plane by epitaxy and can only alter the lattice out of plane. Using  this epitaxial constraint allows one to drive the crystalline lattice (symmetry, rotations, etc...) into distinctly different areas of phase space.  For example, experiments in manganite thin films showed the emergence of a \emph{hidden} phase that existed only in the dynamic realm~\cite{Ichikawa:2011p42146}. Moving into the mid-IR region enables direct pumping of lattice modes that can trigger phase transitions~\cite{Rini:2007p20107} and was recently used to trigger a metal-insulator transition through dynamic strain created by direct pumping of substrate phonons~\cite{Caviglia:2012p39471}. Low energy photons in the THz regime can also serve as a dynamic way to drive transitions with ultra-fast electric fields \cite{Liu:2012p41599}. Such experiments have only begun to explore the complex landscape available in the dynamic realm.

\begin{acknowledgments}
Work at Argonne National Laboratory, including the Advanced Photon Source, is supported by the U.S. Department of Energy, Office of Science under grant no.\ DEAC02-06CH11357. JC was supported by DOD-ARO under grant no.\ 0402-17291. The work in Singapore was supported by the National Research Foundation, Singapore, through Grant NRF-CRP4-2008-04. JMR was supported by ARO (W911NF-12-1-0133) and
acknowledges useful discussions leading to ideas presented in this manuscript  during a workshop sponsored by the Army Research Office
(grant no.\ W911NF-12-1-0171). AJM acknowledges the US Department of Energy, Office of Science, under grant No. DE-FG02-04ER46169.
\end{acknowledgments}

%

\clearpage
\printfigures

\end{document}